\documentclass[aps,pra,twocolumn,10pt]{revtex4-1}
\usepackage{graphicx}

\begin{document}

\title
    [Measurements of the ion velocity]
    {
    Measurements of the ion velocity distribution in an ultracold neutral plasma
    derived from a cold, dense Rydberg gas
    }

\author{S. D. Bergeson$^1$ and M. Lyon$^2$}
\affiliation{$^1$ Department of Physics and Astronomy, Brigham Young University, Provo, UT 84602, USA}
\affiliation{$^2$ Joint Quantum Institute and Department of Physics, University of Maryland, College Park, Maryland 20742, USA}

\begin{abstract}
We report measurements of the ion velocity distribution in an ultracold neutral plasma
derived from a dense, cold Rydberg gas in a MOT. The Rydberg atoms are excited using a resonant
two-step excitation pathway with lasers of 4 ns duration. The plasma forms spontaneously and
rapidly. The rms width of the ion velocity distribution is determined by measuring laser-induced
fluorescence (LIF) of the ions.
The measured excitation efficiency is compared with a Monte-Carlo wavefunction calculation,
and significant differences are observed.
We discuss the conditions for blockaded Rydberg excitation
and the subsequent spatial ordering of Rydberg
atom domains. While the blockade interaction is
greater than the Rabi frequency
in portions of the atomic sample,
no evidence for
spatial ordering is observed.
\end{abstract}


\centerline{Submitted to J. Phys. B Special issue on Rydberg atom physics}
\maketitle

\section{Introduction}

The Rydberg blockade
\cite{PhysRevLett.85.2208, PhysRevLett.87.037901, Comparat:10,
0953-4075-45-11-113001, RevModPhys.82.2313}
occurs when a laser excites one atom to a Rydberg state in the presence of several ground-state atoms.
Under certain conditions, the subsequent interactions between the Rydberg atom and neighboring ground-state atoms shifts the energy levels in such a way that none of the surrounding atoms can be excited by the laser. This blockade effect occurs over a characteristic distance scale given by the details of the atom-atom interaction.

In an atomic sample where the nearest-neighbor distance between atoms is much smaller than the blockade radius, the excitation blockade can lead to spatial ordering of the Rydberg atoms in the gas
\cite{Schauss27032015,PhysRevA.78.040704,PhysRevLett.98.023002,
PhysRevLett.104.013001,PhysRevLett.109.053002,PhysRevLett.110.253003,
PhysRevA.88.061406,PhysRevLett.114.203002}.
A spatially-ordered atomic system has many potential applications, not only in massively parallel quantum logic, but also in probing the behavior of many-body systems, engineering synthetic gauge fields,
enhancing fusion in high-energy-density plasmas, and generating high brightness charged particle beam sources \cite{Comparat:10,0953-4075-45-11-113001,RevModPhys.82.2313, salpeter69,ichimaru93,
PhysRevLett.115.214802}.

The connection between dense Rydberg systems and plasma physics is very close.
A number of experiments have reported that dense Rydberg systems spontaneously convert
into plasmas \cite{vitrant82, bergeson98b, gallagher03, walz04, vincent13, mcquillen13, lyon13, sadeghi14}.
Although the initial atom temperature can be a few mK, the subsequent ion temperature
is typically hundreds of times
higher because of unbalanced forces between the ions
\cite{PhysRevLett.110.253003,lyon11,chen04,lyon13}. This process is called disorder-induced heating (DIH),
and it can severely limit the degree of strong coupling in the plasma.
A plasma generated from
an initially ordered system would circumvent disorder-induced heating, which is the largest source
of heating in ultracold plasmas.
The resulting ultracold system could become a robust platform for transport and kinetic studies near the plasma/solid interface \cite{strickler15}, with interesting applications for warm dense matter experiments \cite{murillo07}.
It may be possible to generate an ordered plasma system from a cold Rydberg gas that is
excited in the Rydberg blockade regime \cite{bijnen11, bannasch13, mcquillen13}.

Achieving spatial order in the laboratory for mm-sized atomic samples requires careful manipulation of the many-body wavefunction. The ground state of the ordered Rydberg crystal has one excited atom surrounded by several atoms in the ground electronic state. As discussed, for example, in Ref. \cite{0953-4075-45-11-113001}, the many-body wavefunction for $N_b$ blockaded atoms has the form $\left|W\right> = (N_b)^{-1/2} \sum_{i=1}^{N_b} \left|g_1, g_2, g_3, ..., e_i, ... g_{N}\right>$. The adiabatic transition to this low-energy and highly-ordered state from one in which all atoms are initially in the electronic ground state requires a shaped chirped pulse \cite{Schauss27032015,bannasch13}. However, even when the excitation pulse is not specifically tailored, blockading still occurs \cite{tong04} and this changes the spatial distribution of excited-state atoms \cite{Schauss27032015}.

Modifications in the initial nearest-neighbor distribution in an ultracold neutral plasma can lead to a significant reduction in the resulting disorder-induced heating
\cite{sadeghi14}. This suggests that straight-forward excitation to Rydberg states using ns-duration pulsed lasers \cite{tong04}, followed by rapid ionization, could lead to an ultracold neutral plasma in which the ion temperature is lower than in a plasma created by direct excitation to the continuum.

In this paper, we explore the possibility of creating an ultracold neutral
plasma with reduced temperature by exciting laser-cooled calcium atoms in a MOT to Rydberg states.
The conditions
under which blockading is expected to occur are presented. The Rydberg atom interactions lead to rapid ionization, as in Ref. \cite{PhysRevLett.100.043002,PhysRevA.90.022712}, as well as ion heating.
We report measurements of the ion velocity distribution in the plasma derived from these
interacting Rydberg atoms. While the blockade interaction is
greater than the Rabi frequency
in portions of the atomic sample, no evidence for blockaded excitation is observed.

\section{Blockade discussion}

\begin{figure}[b]
  \centerline{\includegraphics[width=0.95\columnwidth]{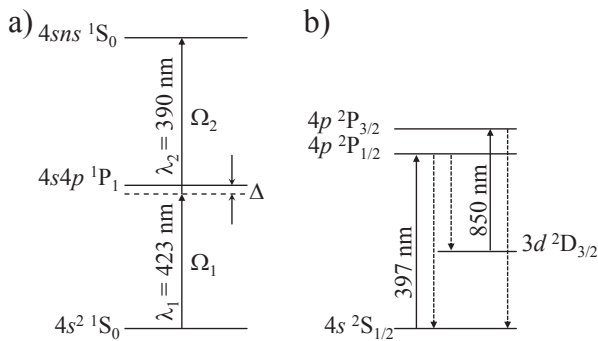}}
  \caption{\label{fig:level1}
  a) Partial energy level diagram of our two-step Rydberg excitation scheme in neutral Ca. Pulsed lasers at 423 and 390 nm drive the transition from the ground state to the Rydberg level.
  b) Partial energy level diagram in Ca$^+$. We use a cw laser at 397 nm to excite the $s-p$ transition and measure fluorescence at this same wavelength. A laser at 850 nm is used to depopulate the metastable state.
  }
\end{figure}

Excellent treatments of the Rydberg blockade have been reported in the literature
\cite{Comparat:10,0953-4075-45-11-113001, RevModPhys.82.2313}. The blockade energy is
\begin{equation}
  U_6 = \frac{C_6}{r^6},
\end{equation}
where $r$ is the distance between ground-state atoms. The $C_6$ coefficients have been calculated
for the $4sns~^1S$ Rydberg series in calcium \cite{0953-4075-45-13-135004}. In atomic units,
the coefficients are calculated using the expression
\begin{equation}
  C_6 = n^{11} \left( an^2 + bn+c \right),
\end{equation}
where $n$ is the principal quantum number of the Rydberg state and $a=-1.793 \times 10^{-3}$, $b=0.3190$, and $c=-1.338$. This expression is valid for $30 \leq n \leq 70$. Conversion from atomic units
is accomplished using the expression $C_6~(\mbox{GHz}\cdot\mu\mbox{m}^6) = 1.4448\times 10^{-19}~(\mbox{a.u.})$. For an atom separation of 2 $\mu$m, excitation to $n=45$ gives an interaction energy of 33 MHz. This energy is a strong function of $n$, so exciting to $n=55$ increases the interaction energy by an order of magnitude. This energy also scales with density-squared.

The time-scale for Rydberg excitation depends on laser intensity, principle quantum number, and density. Similar to most other current Rydberg experiments \cite{mcquillen13,PhysRevA.88.043430,PhysRevA.91.062702}, we use a two-step process to excite Rydberg levels. Our experiment uses laser pulses at 423 and 390 nm (see Fig. \ref{fig:level1}).
The Rabi frequency for the 390 nm excitation, the second step in our experiment, can be calculated if the transition matrix element is known. In calcium it is probably similar to the rubidium $5p\rightarrow ns$ value \cite{RevModPhys.82.2313},
$\left< r \right> = 0.014 \times (50/n)^{3/2} a_0$. The Rabi frequency is calculated as
\begin{equation}
  \left| \Omega_2 \right| = e{\cal{E}}_2 \left< r \right> / \hbar ,
\end{equation}
where $e$ is the electron charge and ${\cal{E}}_2$ is the electric field from the 390 nm laser. For excitation to $n=40$ and a laser intensity of $10^8$ mW/cm$^2$, the Rabi frequency is $\Omega_2 = 2\pi \times 300~\mbox{MHz}$. As discussed in Ref. \cite{0953-4075-45-11-113001}, when there are $N_b^{}$ atoms in a blockade sphere, then the Rabi frequency is increased by a factor of $N_b^{1/2}$ and also by a coordination number that includes the influence of neighbors beyond the nearest one.

The Rabi frequency can also be written in terms of the laser intensity,
\begin{equation}
  \Omega = 2\pi \times \Delta\nu_N^{} \sqrt{\frac{I}{2I_{\rm sat}}},
\end{equation}
where $\Delta\nu_N^{} = (2\pi\tau)^{-1}$ is the natural linewidth, $\tau$ is the excited state lifetime, $I$ is the laser intensity, and $I_{\rm sat}$ is the saturation intensity. A large Rabi frequency power-broadens the atomic transition so that the full-width at half-maximum (FWHM) is,
\begin{equation}
  \delta\nu = 2\pi \times \Delta\nu_N \sqrt{\frac{I}{I_{\rm sat}}} = \sqrt{2}\Omega.
\end{equation}
This power-broadened linewidth sets the time-scale for the blockade.
The assumption in all Rydberg blockade work is that the
interaction energy is much greater than the
Rabi frequency, or $U_6/\hbar \gg \Omega$.

As discussed previously, the interaction energy depends on the nearest-neighbor distance. In
our thermal sample, these distances have a distribution. Consequently, depending on the
excitation parameters, portions of the atomic sample may be in the blockade regime, as
discussed below.

\section{Nearest-neighbor distribution in a MOT}

\begin{figure}[t]
  \centerline{\includegraphics[width=0.9\columnwidth]{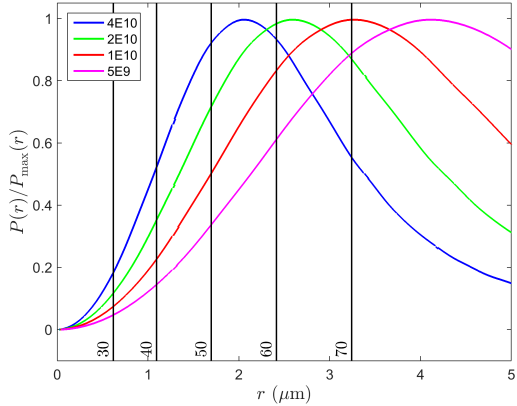}}
  \caption{\label{fig:dist}
  A plot of the nearest-neighbor distribution as a function distance between atoms, $r$, for a range of peak densities, $n_0$. The vertical lines indicate the nearest-neighbor distances at which the $U_6$ energy is equal to 300 MHz. The relative number of blockaded atoms can be determined by integrating the distribution up to the blockade radius.
  }
\end{figure}

For a thermal gas at constant density, the nearest-neighbor distribution is given by the Erlang distribution,
\begin{equation}
  P_1(r) = 3\left(\frac{r}{a_{ws}^{}}\right)^2 \exp\left(-\frac{r^3}{a_{ws}^3}\right),
\end{equation}
where $a_{ws}^{} = (3/4\pi n_0^{})^{1/3}$ is the Wigner-Seitz radius and $n_0^{}$ is the density. In a MOT, where the density is Gaussian of the form $n(r) = n_0^{} \exp(-r^2/2\sigma^2)$, the nearest neighbor distribution changes. At small values of $r$, it has nearly the same form as $P_1(r)$. However, the distribution falls off much more slowly when $r>\sigma$ because the density falls off gradually and the average nearest-neighbor distance increases.

In ultracold neutral plasmas, most of the disorder-induced heating (DIH) comes from ions that are much closer to each other than the average distance. If the nearest-neighbor distribution could be suppressed at short distances, the DIH temperature would be reduced \cite{sadeghi14}. If the initial order matched the close-packed configuration of a strongly-coupled plasma, DIH would be eliminated \cite{lyon15b}.

Depending on the experimental conditions, portions of the atomic cloud are found to be in a
regime in which blockaded excitation is expected to occur.
We derive the nearest-neighbor distribution in the MOT by generating a three-dimensional random-normal distribution and computing the distance from each atom to its first nearest neighbor.
In Fig. \ref{fig:dist} we plot this distribution for a few different peak MOT densities.
Also shown in Fig. \ref{fig:dist} are vertical lines that indicate the radius at which
the $U_6$ interaction energy is equal to 300 MHz for a range of principle quantum numbers.
This value is chosen because it matches the measured frequency width (FWHM) of our laser pulses.
Because the interaction energy scales as the nearest-neighbor distance to the sixth power,
atoms to the left of the vertical lines should be
strongly within the blockaded excitation regime.
We therefore expect the leading edge of the nearest-neighbor distribution
to be suppressed in these systems \cite{sadeghi14}.

\section{Experimental details}

Calcium atoms in a thermal beam (T = 750 K) are slowed and captured into a MOT (T $\sim$ 0.001 K). The Rydberg atoms are excited using a two-step process, as shown in Fig. \ref{fig:level1}. The first step is nearly-resonant with the 423 nm laser-cooling transition, with $\Delta = 2\pi \times 150~\mbox{MHz}$. A cw laser beam at 423 nm is pulse-amplified in a single transverse-pumped dye cell.
An AOM is used to turn the cw laser beam on 100 ns before the pump laser arrives at the dye cell.
The pump laser pulse width is 4 ns FWHM. The pulse-amplified 423 nm laser beam is expanded, apertured to a few mm diameter, attenuated, and imaged into the MOT. The resulting intensity profile is relatively flat, with $I/I_{\rm sat} = 100$ so that the power-broadened linewidth is 350 MHz, approximately equal to the bandwidth of the laser. The corresponding
Rabi frequency for this transition is $\Omega_1^{} = 2\pi \times 250~\mbox{MHz}$.

The $4p-ns$ transition is driven using a pulse-amplified frequency-doubled ti:sapphire laser near 390 nm. An AOM is also used in this laser beam to turn the cw laser on 100 ns before the pump laser arrives at
the dye cell. As with the 423 nm laser, this laser is also apertured and imaged into the MOT so that the intensity is in the range of $10^9$ mW/cm$^2$. The two lasers counterpropagate each other in the MOT and arrive simultaneously. The intensity of this laser is adjusted so that the Rabi frequency ranges from 1 to 300 MHz, similar to other
ns-pulsed experiments \cite{PhysRevLett.107.243001,kubler10}.

Laser pulses with durations in the $\sim 1~\mu$s range are typically used in blockade experiments. Unfortunately for us, the atoms in our MOT move on the $\mu$s time scale. At a typical density of $3\times 10^{10}~\mbox{cm}^{-3}$ and a temperature of 1 mK, the average time between nearest-neighbor collisions is only 2 $\mu$s. This motion results in collisions, placing $\mu$s-duration pulsed laser experiments in Ca outside of the ``frozen gas'' assumption used in all blockade experiments (see Fig. \ref{fig:level1}).
Unless the atoms are much colder or the atoms tightly confined
to short distances, ns-duration laser pulses are required.

Rydberg excitation efficiency is measured using MOT loss measurements. A weak, resonant cw 423 nm laser beam passes through the MOT. When the MOT is loaded, the transmission of this laser beam is typically 1\%. When the MOT atoms are excited to Rydberg states, they no longer interact with this laser beam and the transmission increases. Using Beer's law we are able to convert this transmission into atomic density. In a typical measurement, we measure the changes in transmission of this probe laser beam as the 390 nm laser frequency is scanned across the Rydberg excitation resonance. To avoid EIT effects, the probe laser is turned off while the excitation lasers are on.

\begin{figure}[b]
  \centerline{\includegraphics[width=0.9\columnwidth]{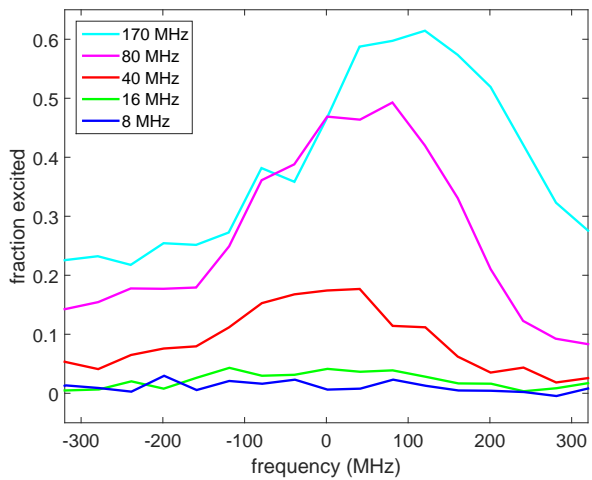}}
  \caption{\label{fig:scan1}
  Measurements of the fraction of MOT atoms excited to the $4s30s~^1S_0^{}$ state vs. the 390 nm laser frequency for a number of different laser intensities. The peak density in the MOT is $n_0 = 4\times 10^{10}~\mbox{cm}^{-3}$. The excited fraction is calculate based on the transmission of a weak probe laser
  beam that passes through the MOT, in resonance with the 423 nm MOT transition.
  As the 390 nm Rydberg excitation laser intensity increases, a greater number of atoms are excited into the Rydberg state, saturating at roughly 50\% excitation. Increasing intensity also Stark shifts and broadens the transition. The legend indicates the Rabi frequency, $\Omega_2 / 2\pi$. For these measurements, $\Omega_1^{} = 2\pi \times 250~\mbox{MHz}$.
  }
\end{figure}

The atomic spectroscopy is performed by offset-locking the lasers to a partially-stabilized frequency comb \cite{lyon2014}. This provides long-term stability, accurate determinations of the absolute frequencies, and convenient frequency scanning of the lasers.
The comb uses a passively mode-locked ti:sapphire laser with a repetition rate, $f_{\rm rep}^{} = 978 \mbox{MHz}$. Its spectrum is nominally Gaussian, 35 nm FWHM, and centered at 815 nm. Without broadening, the spectrum gives strong beat notes with all of the cw lasers used in this experiment ($>25$ dB above the technical noise, 30 kHz measurement bandwidth).
One mode of the comb is offset-locked to a 780 nm diode laser that is itself locked to the $^{87}\mbox{Rb}~F=2\rightarrow F=2/3$ cross-over transition. The offset lock feeds back to the ti:sapphire laser cavity length. The absolute frequency of one comb mode is therefore known with a precision of a few kHz. The repetition rate of the laser is not controlled. However repeated measurements of the repetition rate in a 1-second measurement duration show drifts of approximately 1 Hz/s. Therefore, counting the repetition rate during our experiments allows us to determine laser frequencies with sub-MHz precision.

\section{Pulsed excitation measurements}

\begin{figure}
  \centerline{\includegraphics[width=0.9\columnwidth]{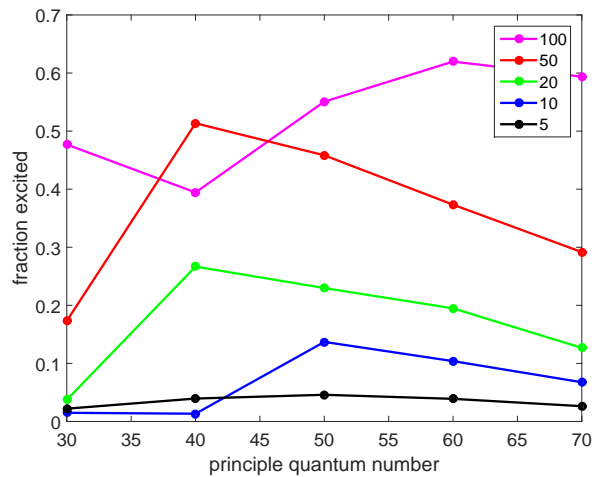}}
  \caption{
  \label{fig:analysis}
  The fraction of MOT atoms excited to Rydberg states as a function of the principle quantum number. The legend indicates the peak intensity in the 390 nm laser pulse in units of $10^7~\mbox{W/cm}^2$.
  }
\end{figure}

In Fig. \ref{fig:scan1} we show the fraction of MOT atoms excited to the $4s30s~^1S_0^{}$ state as a function of the 390 nm laser frequency. The data shows increasing excitation as $\Omega_2$ increases.
For excitation to this level, we expect no blockading to occur. The peak density in the MOT is $n_0 = 4\times 10^{10}~\mbox{cm}^{-3}$. As the $\Omega_2$ increases, the excitation fraction grows, and the line center is Stark shifted and broadened. The signal-to-noise ratios in these data are compromised by shot-to-shot fluctuations in the excitation lasers, density fluctuations in the MOT, and by technical noise in the strongly-attenuated probe laser beam.

We have collected data like this for $n=30,40,50,60,$ and $70$. As the excitation proceeds to greater values of $n$, the lines are significantly broadened even at low excitation fractions. At $n=70$, for example, the line is broadened beyond 800 MHz.

In Fig. \ref{fig:analysis} we plot the peak excitation fraction as a function of principle quantum number for a range of different 390 nm laser powers. For the highest laser powers, the excitation fraction never varies far from 0.5. This indicates that there is relatively little ionization or other loss on the time scale of the Rydberg excitation pulse. The highest laser power probably power-broadens the Rydberg transition so that it exceeds the $U_6$ energy. At higher values of $n$, the $ns$ and $(n-1)d$ states are only a few hundred MHz apart, and the transitions are not well separated.

\begin{figure}
  \centerline{\includegraphics[width=0.9\columnwidth]{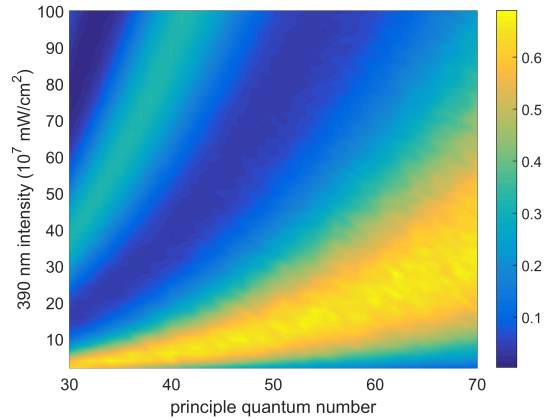}}
  \caption{\label{fig:exfrac}
  The calculated excitation fraction in the Rydberg state after the laser pulses
  are over for a range of values of the
  principle quantum number (30 to 70) and a range of 390 nm laser intensities
  (2 to $100 \times 10^7$ mW/cm$^2$). The greatest
  excitation occurs when the lasers produce a single $\pi$-rotation
  in the wavefunction. This calculation was performed on resonance ($\Delta=0$)
  and averaged over 1000 runs (see Eq. \ref{eqn:mcwf}).
  }
\end{figure}

Two features stand out in this figure. We will consider the red trace for a laser intensity of $50\times  10^7~\mbox{W/cm}^2$. The excitation fraction increases with increasing $n$ from 30 to 40. At higher
values of the principle quantum number, the excitation fraction falls slowly.
This behavior is also observed for most of the other intensities, although the value of $n$
above which the excitation fraction falls appears to be greater at lower intensities.

\section{Monte-Carlo wavefunction calculation}

To understand the excitation process, as well as the possible influence of blockading on the excitation dynamics, we have constructed
a three-level model based on the Monte-Carlo wave function (MCWF) method \cite{PhysRevLett.68.580,Molmer:93}.
This method is appropriate for calculating excitation amplitudes in the presence of radiative
decay.
We begin with a single-atom wavefunction
$
\left| \psi \right> = a(t) \left|g \right> + b(t) \left|e \right> + c(t) \left|r \right>
$
where $\left|g\right>,\left|e\right>,$ and $\left|r\right>$ represent the $4s^1~^1S_0^{}$, $4s4p~^1P_1^{\circ}$, and $4sns~^1S_0^{}$
levels and $a(t), b(t),$ and $c(t)$ are their time-dependent amplitudes.
The Schr\"{o}dinger equation is used with a rotating-wave approximation to find the time-evolution
of the eigenstate amplitudes.
In the $n^{\rm th}$ time
interval $dt$, these amplitudes change according to
\begin{eqnarray}
  dc_n & = b_{n-1} \frac{i \Omega_{2}}{2} dt \nonumber \\
  db_n & = a_{n-1} \frac{i \Omega_{1}}{2} dt + b_{n-1} \left( -\frac{\Gamma}{2} - i\Delta \right) dt + c_{n-1} \frac{i \Omega_{2}}{2} dt \nonumber \\
  da_n & = b_{n-1} \frac{i \Omega_{1}}{2} dt ,
  \label{eqn:mcwf}
\end{eqnarray}
where $\Gamma = 1/\tau = (4.5~\mbox{ns})^{-1}$ \cite{PhysRevA.64.012508,PhysRevA.67.043408}.
This is equivalent to treating both transitions in a rotating frame, with each
frame rotating independently of the other. The Rydberg state is assumed to be stable.

Decay of the intermediate $\left|e\right>$ state is calculated probabilistically.
In the standard MCWF manner, we calculate a probability
$dp_n = \left| b_n \right| \Gamma dt$ and compare it
to a random number $\epsilon$ generated with uniform probability between 0 and 1.
If $\epsilon < dp_n$, then the state $\left|e\right>$ has decayed and we set
the amplitudes $a, b,$ and $c$ to $1, 0,$ and $0$ and proceed with the calculation.
To ensure proper normalization, we calculate
$(\left|a_n \right|^2 + \left|b_n \right|^2 + \left|c_n \right|^2)^{1/2}$ at
every time step and divide this into the amplitudes. Because
$dt \ll \Omega_{423}^{-1}, \Omega_{390}^{-1}$, this amplitude correction is small
and is equivalent to the normalization method used in Refs. \cite{PhysRevLett.68.580,Molmer:93}.

The calculated value of $\left| c(t_{\infty}) \right|^2$ is plotted in Fig. \ref{fig:exfrac}.
This corresponds to the probability of finding an
atom in the Rydberg state after the laser
pulses are over
As expected, the simulation shows the population in the Rydberg state oscillating
at the frequency $\textstyle{\frac{1}{2}}\left(\Omega_1^2 + \Omega_2^2\right)^{1/2}$ during
the laser pulse.
The greatest excitation
occurs when a the pulse intensity and dipole moment produce a single $\pi$-rotation.

The simulation in Fig. \ref{fig:exfrac} qualitatively reproduces
some features observed in the experiment.
For example, it captures the low-amplitude excitation
for lower values of the principle quantum number, and increasing excitation with $n$ at a given
laser intensity.
However, the profiles in Fig. \ref{fig:analysis}, which correspond to horizontal
cuts across the data in Fig. \ref{fig:exfrac}, depend differently on laser
intensity. The experimental data of Fig. \ref{fig:analysis} shows that excitation
decreases with lower laser power. The simulation data of Fig. \ref{fig:exfrac}
shows that excitation \textit{increases} with decreasing laser intensity. In the simulation,
this happens because the higher intensities produce multiple Rabi oscillations, resulting
in lower overall excitation after the laser pulses are over.

These differences may be partly due to collective excitation and interactions of the Rydberg levels,
effects that are omitted from the simulation.
However, other factors may also be important. In the experiment, we observe significant shot-to-shot
variation in the laser power. The pulsed laser amplifiers use a multi-longitudinal-mode Nd:YAG
laser. The fast temporal structure in these pulses can produce broad wings in the spectral profile,
whereas the simulation is performed on resonance for a single frequency. Finally, interactions between
the Rydberg atoms beyond collective excitation could be important.

\section{Ion velocity distribution measurements}

\begin{figure}
  \includegraphics[width=0.95\columnwidth]{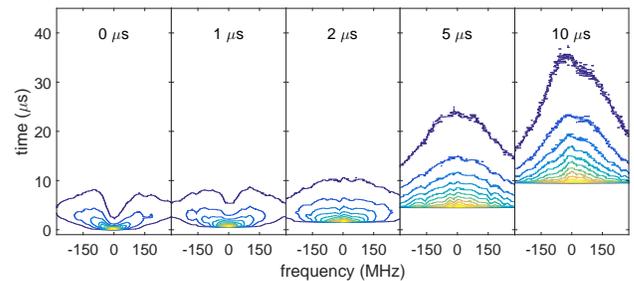}
  \caption{\label{fig:delaytime}
  A countour plot of the laser-induced fluorescence signal. The atoms are excited into
  the state $n=50$. After a delay time ranging from 0 to 10 $\mu$s, the 397 nm probe laser
  is rapidly switched on. For each of these plots, the laser-induced fluorescence is
  measured as a function of time at a given 397 nm laser frequency. Measurements are
  repeated for a range of 397 nm laser detunings. A collection of measurements are then
  used to produce these countour plots. The 850 nm repump laser is used to prevent
  atoms from accumulating in the metastable state.
  }
\end{figure}

After the Rydberg atoms are created, they spontaneously ionize.
A recent study in Ref. \cite{PhysRevLett.100.043002} proposed a many-atom ionization mechanism.
In a dense Rydberg gas, the configuration with all individual atoms in the well-defined Rydberg
state is not necessarily an eigenstate of the many-body system. The authors of Ref. \cite{PhysRevLett.100.043002} suggest that the excited Rydberg atoms make a series of transitions to nearby excited states and
that some of these atoms eventually end up in the continuum.
This state diffusion is predicted to be the major ionization mechanism when $n=50$ at a density of $10^{11}~\mbox{cm}^{-3}$. If the Rydberg atoms transition from the initial
Rydberg state to nearby states that are dipole-coupled, the mechanical force on the atoms can
collisionally ionize the atoms in the 100 ns time scale noted in their experiment. A recent simulation of
the experiment could not reproduce the prompt ionization of the experiment
\cite{PhysRevA.90.022712}. However, electron collisions
and collisions with fast atoms were not included in the simulation. Those effects
may boost the ionization signal at early times.

A calculation in Ref. \cite{0953-4075-44-18-184015} suggests that the avalanche ionization
rate per atom at a density of $4\times 10^{10}~\mbox{cm}^{-3}$ and a principle quantum number of
50 is only $10^4~\mbox{s}^{-1}$. This is far too slow to explain ionization. Direct collisional
ionization of the Rydberg atoms due to thermal motion occurs on the $\sim1~\mu\mbox{s}$ time scale
\cite{0953-4075-38-2-024}.

The mechanical forces between Rydberg atoms has been documented in a number of publications.
This happens when some mechanism transfers atoms from the initial Rydberg state to some
other state. This can occur through collisions \cite{PhysRevLett.94.173001}, interactions
at F\"{o}rster resonances or collisional resonances \cite{PhysRevA.78.040704, Comparat:10,PhysRevLett.109.053002,PhysRevLett.94.173001},
through blackbody radiation \cite{PhysRevLett.98.023004},
or transitions induced using microwaves
\cite{0953-4075-44-18-184020,PhysRevLett.115.013001,PhysRevA.84.052708}. When present, these mechanical forces will
accelerate Rydberg atoms towards (or away from) one another. If this Rydberg gas is converted into
an ultracold plasma, the ions will inherit the velocity distribution of the Rydberg atoms.

In our experiment, similar to Ref. \cite{PhysRevLett.100.043002}, we observe a rapid rise in
the ion signal following excitation to Rydberg states. However, complete conversion of the
Rydberg system into a plasma seems to occur over a few microseconds.

Using laser spectroscopy, we probe the Rydberg system after it has been excited. We use a
cw laser at 397 nm to drive the $4s-4p$ transition, as in our previous ultracold neutral plasma
experiments \cite{lyon15}.
When the Rydberg atoms are in states
with low angular momentum, this ``inner core'' excitation leads to extremely fast ionization.
A recent study on similar states in Sr showed that the width of this
auto-ionizing resonance is measured in THz \cite{0953-4075-44-18-184001},
suggesting that autoionization occurs in less than one ps.
The Rydberg atom is instantly converted into an ion
when the inner electron is excited. That ion initially
has the same velocity as the Rydberg atom. The combined motion of
the Rydberg atoms before ionization and also the motion of the ions after ionization
both contribute to the width of the laser-induced fluorescence signal at 397 nm.

In Fig. \ref{fig:delaytime} we present contour plots of the
ion fluorescence signal as a function
of frequency and time. In these measurements we introduce a delay between the time when the
Rydberg states are initially populated and when the 397 nm probe laser turns on. This
provides insight into the time-scale for transferring atoms out of the initial
Rydberg state by collisions or other mechanisms. For each of the measurements, a
repump laser at 850 nm is used to prevent population from accumulating in the metastable
$d$-state (see Fig. \ref{fig:level1}).

\begin{figure}
  \includegraphics[width=0.95\columnwidth]{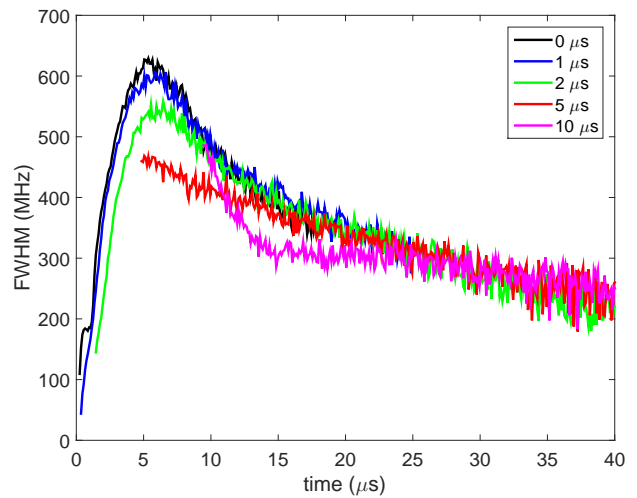}
  \caption{\label{fig:fwhmplot}
  The measured FWHM of the data from Fig. \ref{fig:delaytime}. The lineshapes
  are complicated and asymmetric. However, the FWHM gives some indication of the
  approximate velocity distribution. Atomic interactions give rise to these
  large widths. The contour lines correspond to 90\%, 80\%, etc., levels of the maximum fluorescence signal in the figure.
  }
\end{figure}

For time delays of 0 and 1 $\mu$s, the fluorescence is depleted when the 397 nm laser is on
resonance. In addition, the entire fluorescence signal vanishes in less than 10 $\mu$s.
For these measurements, autoionization of Rydberg atoms
is the dominant loss mechanism when the laser is on resonance. At later
delay times, the fluorescence signal is less strongly perturbed by the 397 nm laser.
At these times, the Rydberg atoms have been converted into ions or
diffused into high angular momentum states in which the 397 nm probe laser beam
does not significantly perturb the velocity distribution.

In Fig. \ref{fig:fwhmplot} we measure the FWHM as a function
of time for all of the data in Fig. \ref{fig:delaytime}. While it could be argued that
the 0 and 1 $\mu$s data are strongly perturbed by the 397 nm laser, it appears that the longer
delay data are not. The width of the fluorescence signal rises to several
hundred MHz, corresponding to velocities on the order of 200 m/s. This occurs with
and without the presence of the 397 nm laser. The nearly linear rise
in the fluorescence width indicates that the ion acceleration is approximately $4\times 10^7$ m/s$^2$.

The central question in this investigation is whether or not the excitation of the Rydberg atoms
is blockaded by the Rydberg-Rydberg interaction, and
if that blockaded excitation would lead to a spatial ordered
Rydberg system. The primary manifestation of that blockading
and ordering
should be the slow expansion of the plasma. Data previously published by us
on the expansion of ultracold neutral plasmas
shows characteristic post-DIH broadening of the velocity distribution
of roughly $2\times 10^7$ m/s$^2$ \cite{lyon13}, depending on electron temperature.
This value is same order of magnitude as the value derived from Fig. \ref{fig:fwhmplot}.
We therefore must conclude that the Rydberg blockade is likely not
significant in these data.

It is possible that blockading and ordering is present but that it is masked by some other effect.
The plasma expansion is driven by the electrons. Anything that heats the electrons
will contribute to faster plasma expansion.
Energy-pooling collisions between free electrons and excited-state ions and
auto-ionization from the excited inner-shell electron, for example,
could contribute significantly to electron heating.

\section{Conclusion}

We have explored evidences of blockaded Rydberg excitation of laser-cooled
Ca atoms in a MOT
using ns-duration laser pulses. The energy scales set by the
Rabi frequency and the $U_6$ interaction seem to be appropriate for blockaded excitation and spatial ordering
in portions of the atom sample. However, our velocity measurements
provide no indication that this effect is present.

The presence of charged particles in our MOT would prevent realization of the blockade.
It is possible that the transition from a Rydberg atom to an ion occurs on time scales shorter
than the 4 ns laser pulse. Under certain conditions, we measure ions emitted from our MOT.
We have minimized these in the present work by applying a constant electric field until 100 ns
before the laser pulse arrives, effectively sweeping all charged particles from the MOT
when the Rydberg excitation occurs.

It is also possible that blockaded excitation occurs, but that the initial Rydberg state is
rapidly scrambled by interactions with other particles. This would lead to large mechanical forces,
as mentioned previously, that could produce ions rapidly.

It would be better to excite these Rydberg atoms using a longer laser pulse, perhaps 50 ns in duration
with a smooth temporal profile. Such an excitation pulse would place less stringent constraints
on the Rabi frequency. One could even imagine tailoring the frequency and intensity of
such a laser pulse. However, all laser issues aside, the uncontrolled nearest-neighbor distribution
in a MOT would undoubtedly complicate the blockade mechanism, producing an uncontrolled
number of Rabi cycles and resulting
in low excitation probability (see Fig. \ref{fig:exfrac}) and weak blockading.

\section{Acknowledgements}

This research is supported in part by the Air Force Office of Scientific Research (Grant No. FA9950-12-0308)
and by the National Science Foundation (Grant No. PHY-1404488).


\providecommand{\newblock}{}

\end{document}